\documentstyle[preprint,eqsecnum,aps]{revtex}

\begin{document}
\draft
\baselineskip=24pt plus .5pt minus .2pt

\title{Semi-classical Characters and Optical Model Description of \\
Heavy Ion Scattering, Direct Reactions, and Fusion \\
at Near-barrier Energies}

\author{B. T. Kim, W. Y. So, and S. W. Hong}
\address{Department of Physics and Institute of Basic Science, 
Sungkyunkwan University, Suwon 440-746,
Korea}\
\author{T. Udagawa}
\address{Department of Physics, University of Texas, Austin, Texas 78712}\
\date{\today}
\maketitle
\begin{abstract}
An approach is proposed to calculate the direct reaction (DR) and fusion 
probabilities for heavy ion collisions at near-Coulomb-barrier energies as 
functions of the distance of closest approach $D$ within the framework of the
optical model that introduces two types of imaginary potentials, DR and fusion. 
The probabilities are calculated by using partial DR and fusion cross sections, 
together with the classical relations associated with the Coulomb trajectory. 
Such an approach makes it possible to analyze the 
data for angular distributions 
of the inclusive DR cross section, facilitating the determination of the radius 
parameters of the imaginary DR  potential in a less ambiguous manner. 
Simultaneous $\chi^{2}$-analyses are performed of relevant data for the 
$^{16}$O+$^{208}$Pb system near the Coulomb-barrier 
energy.
\end{abstract}

\vspace{2ex}


\narrowtext

\section{Introduction}    

 Collisions between heavy ions at near-Coulomb-barrier energies are very much
governed by the Coulomb potential involved, and thus the general features of 
the elastic scattering and direct reaction (DR) data can be understood based 
on the idea that the colliding ions primarily move along a classical 
Coulomb trajectory~\cite{bass}. 
These features are seen most dramatically in plots of the ratios of 
the elastic differential cross section ($d\sigma_{E}/d\Omega$) 
or the inclusive (sum of all different) DR one ($d\sigma_{D}/d\Omega$) 
to the Rutherford differential 
cross section ($d\sigma_{c}/d\Omega$), {\it i.e.}, 
\begin{equation}
P_{i} \equiv \frac{d\sigma_{i}}{d\Omega}/\frac{d\sigma_{c}}{d\Omega} 
              \;(=\frac{d\sigma_{i}}{d\sigma_{c}}), \;\;\;\;\;
              (i= E~ \mbox{or}~ D),
\end{equation}
as a function of the distance of closest approach $D$ 
(or the reduced distance $d$)~\cite{bass,sat1} 
that is related to the scattering angle $\theta$ through
\begin{equation}
D =d(A_{1}^{1/3}+A_{2}^{1/3})
=\frac{1}{2}D_{0} \left( 1+\frac{1}{\mbox{sin}(\theta/2)} \right) 
\;\;\;\;\;\;\;\;\;\;  \mbox{with} \;\;\;\;\;\;\;\;\;\; 
D_{0}=\frac{Z_{1}Z_{2}e^{2}}{E_{cm}}.
\end{equation}
Here $D_{0}$ is the distance of closest approach in a head-on collision 
($s$ wave). Further, ($A_{1}, Z_{1}$) and ($A_{2}, Z_{2}$) are the mass and 
charge of the projectile and target ions,  respectively, and
$E_{cm}$ ($E_{lab}$) is the incident energy in the center of mass (laboratory) 
system. 
$P_{E}$ and $P_{D}$ thus defined may be called the elastic and DR 
probabilities, respectively.  
The impact parameter $b$ and orbital angular momentum $\ell$, 
specifying the trajectory, are related to $\theta$ and $D$ by
\begin{equation}
b=\frac{\ell}{k}=\frac{D_{0}}{2}\mbox{cot}\frac{\theta}{2}=
    \sqrt{D(D - D_{0})},
\end{equation}
where $k$ is the wave number.

 As an illustration, we show in Fig.~1 such plots of $P_{E}$ and $P_{D}$ 
for the $^{16}$O+$^{208}$Pb system~\cite{vide,valg} 
at five different incident energies of $E_{lab}$=80, 84, 90, 96, and 102 MeV 
for $P_{E}$ and at  a single energy of $E_{lab}$=90~MeV for $P_{D}$, 
where the data are available. 
As seen, the values of $P_{E}$ at different energies line up to 
form a very narrow band and take a value very close to unity for, say,
$d > 1.65$~fm ($\equiv d_{I}$, interaction distance). 
When $d$ becomes smaller than $d_{I}$, $P_{E}$ falls 
off very rapidly, approximately exponentially.

 The observed behaviour of $P_{E}$ may easily be explained based on
the physical picture that the projectile ion moves primarily 
along a Coulomb trajectory. 
For the case $d >d_{I}$, the trajectory is far away from the target 
and the projectile is scattered  at the Coulomb scattering angle $\theta$
without being influenced at all by the nuclear force. 
The resultant scattering cross section is thus equal to  
the Rutherford cross section, and $P_{E}$ becomes unity. 
When $d$ becomes smaller than $d_{I}$, however, the incident ion gets  
under the influence of the strong nuclear interaction, and  
absorption takes place, reducing the $P_{E}$-value below unity.
                     
 In accordance with the observed behaviour of $P_{E}$,  $P_{D}$ starts to 
have a significant value at $d \approx d_{I}$. 
It reaches its maximum value ($P_{D} \approx 0.24$) at $d \approx 1.58$~fm, 
where $P_{E}$ becomes approximately 0.7. 
In the region of $d=1.58 \sim 1.65$~fm, the sum $P_{E}+P_{D}$ thus stays 
close to unity. 
This indicates that in that region, the main cause of absorption 
in the elastic channel is DR processes. 
When $d$ becomes still smaller, the sum $P_{E}+P_{D}$ falls off 
rapidly from unity, showing that absorption due to more complicated processes 
eventually leading to fusion takes  place. 
It is remarkable that the sum $P_{E}+P_{D}$ becomes extremely small, say, 
$10^{-3}$ and thus essentially zero, at around $d=d_{c}=1.30$~fm, 
which is the distance for the $s$-wave Coulomb-barrier top.  
This means that the incident flux is almost completely absorbed 
when it reaches that distance. 
It is worth noting that the same picture holds irrespective of the incident 
energy, so long as it is not far away from the Coulomb-barrier energy. 
                              
 Theoretically, we have a very well established optical model 
for evaluating $P_{E}$. This is not the case for $P_{D}$. 
There are a variety of theoretical methods that has been 
proposed for calculating contributions 
from inelastic scattering and transfer reaction processes to $P_{D}$ 
by means of either semi-classical or classical approximations~\cite{bass}. 
It is, however, still a formidable task to carry out calculations  
including all possible processes to obtain a theoretical value of $P_{D}$. 
The aim of the present work is to propose a simple approach to calculate 
$P_{D}$ 
within the framework of an optical model that introduces two types of 
the imaginary potentials; one for DR, the other fusion~\cite{hong,uda1,kim1}. 
We propose to evaluate $P_{D}$ from the partial absorptive 
DR cross sections generated from the optical model calculation.
The underlying assumption is that even after the reaction (removal from
the elastic channel) the projectile ion still moves along the Coulomb 
trajectory, being eventually emitted at the Coulomb scattering angle. 
Under the assumption, we may use the classical relation Eq.~(1.3) to convert 
the partial wave cross section to $P_{D}$. 
                               
The same prescription may also apply to the partial fusion cross sections.
This enables us to evaluate the hypothetical fusion probability $P_{F}$
just as we calculate $P_{D}$. 
The conventional wisdom assumes~\cite{sat1} that
\begin{equation}
P_{E}+P_{D}+P_{F} \approx 1,
\end{equation}
which expresses a simple physical idea that what is absorbed in the
elastic channel at $r=D$ goes into either DR or fusion channels.
Since there is no such measured $P_{F}$-value available, 
it has been impossible to test the above relation experimentally. 
However, it is possible to examine the relation 
by using the theoretical $P_{F}$.
In the present work, theoretical expressions for $P_{D}$ and $P_{F}$ are 
derived in Sec.~II, where we also perform numerical calculations of 
$P_{D}$ and $P_{F}$ along with $P_{E}$ and 
examine the validity of the relation Eq.~(1.4). 
The feasibility of evaluating  $P_{D}$ enables us to analyze 
the angular distribution of the inclusive DR cross section.
This facilitates the determination of the DR part of the optical potential 
with less ambiguity. 
We shall demonstrate this also in Sec.~II.
In Sec.~III, we repeat the simultaneous $\chi^{2}$-analyses that we made 
several years ago~\cite{kim1} for the 
data on the $^{16}$O+$^{208}$Pb system shown in Fig.~1. 
The reanalyses are needed since the fusion data have been revised~\cite{mort} 
after Ref.~\cite{kim1} was published. 
Sec.~IV will then be devoted to our conclusions.

\section{Optical Model Calculations of $P_{D}$ and $P_{F}$ }

\subsection{Derivation of Theoretical Expressions for $P_{D}$ and $P_{F}$ }

 In this subsection, we try to derive theoretical expressions 
for $P_{D}$ and $P_{F}$ within the framework of the optical model. 
We follow the approach proposed some time ago 
to calculate the total DR and fusion cross sections within the optical model
by using imaginary, surface type DR and volume type fusion, potentials,
$W_{D}(r)$ and $W_{F}(r)$, respectively~\cite{hong,uda1,kim1}. 
The total DR and fusion cross sections are then calculated as~\cite{uda2,huss}
\begin {equation}
\sigma_{i} = \frac {2}{\hbar v}
<\chi \raisebox{1ex}{(+)}|W_{\it i}(r)|\chi \raisebox{1ex}{(+)}>
\hspace{.5in} (i=D\;\mbox{or}\;F),
\end{equation}
where $\chi^{(+)}$ is the usual distorted wave function that satisfies
the Schr\"{o}dinger equation with the full optical model potential $U(r)$. 
$\sigma_{D}$ and $\sigma_{F}$ are thus calculated within the same framework as 
the differential elastic scattering  cross section, 
$d\sigma_{E}/d\Omega$, is calculated.
Such a unified description enables us to describe all different types 
of reactions on the same footing.
                           
 The basic ingredients for obtaining theoretical expressions for $P_{D}$ 
and $P_{F}$ are the partial wave cross sections, $\sigma_{i;\ell}$ 
($i= D\;\mbox{or}\;F$), which are obtained by simply expanding the 
cross sections, Eq. (2.1), into the partial wave components;
$\sigma_{i}=\sum_{\ell} \sigma_{i;\ell}$. $\sigma_{i;\ell}$ can explicitly 
be given as~\cite{uda2}
\begin{equation}
\sigma_{i;\ell}=\frac{\pi}{k^{2}}(2\ell +1)T_{i;\ell} 
\hspace{.5in} (i=D\;\mbox{or}\;F),
\end{equation}
where 
\begin{equation}
T_{i;\ell}=\frac{4}{\hbar v} 
                 \int_{0}^{\infty} |\chi_{\ell}(r)|^{2}W_{i}(r)dr.
\end{equation}
In the above equation, $\chi_{\ell}(r)$ is the partial distorted wave function
and $v$ is the relative velocity.

 Eq.~(2.2) with  Eq.~(2.3) is still a quantum mechanical expression, 
where $\ell$ takes only integer values. In what follows, we introduce a few
semi-classical approximations customarily used~\cite{bass,sat1,mott,ford}.
The first is to treat $\ell$ as a continuous variable and to assume that
\begin{equation}
\frac{d\sigma_{i}(\ell)}{d\ell} =\sigma_{i;\ell}.
\end{equation}
We then use the classical relation Eq.~(1.3) that relates $\ell$ to $\theta$. 
It is then straightforward to get
\begin{equation}
\frac{d\sigma_{i}(\ell)}{d\Omega} = 
   \frac{1}{2\pi \mbox{sin}\theta}\frac{d\ell}{ d\theta}
   \frac{d\sigma_{i}(\ell)}{d\ell}
  =\frac{kD_{0}}{16\pi} \frac{1}{\mbox{cos}(\theta/2)\mbox{sin}^{3}(\theta/2)}
   \sigma_{i;\ell}.
\end{equation}
Inserting further Eq.(2.2) into Eq.(2.5), and dividing the resultant expression
by the Rutherford cross section $\sigma_{c}$,  one finally obtains
\begin{equation}
P_{i}=\frac{2\ell +1}{kD_{0}} \mbox{tan}(\theta/2) T_{i;\ell}
                     \approx  T_{i;\ell}
\end{equation}
In obtaining the last expression, use is made of the approximation that
$2\ell + 1 \approx 2\ell$.

 In order that $P_{i}$ can be a probability, it should satisfy 
\begin{equation}
P_{i} \leq 1.
\end{equation}
This requirement is indeed satisfied; in fact we have
\begin{equation}
P_{D}+P_{F}=T_{D;\ell}+T_{F;\ell} \equiv T_{\ell} = 1-|S_{\ell}|^{2},
\end{equation}
where $T_{\ell}$ is the transmission factor and 
$S_{\ell}$ is the partial wave $S$-matrix. 
Since both $P_{D}$ and $P_{F}$ are positive quantities,
it is clear from the above relation that $P_{D}$ and $P_{F}$ should
be less than unity. 
(Note that there is no reason that $P_{E}$ should be smaller than unity. 
Quantum effects such as interference and diffraction may cause the value 
to be greater than unity.)  
Now, for very small $\ell$-values, we expect that $P_{D} \rightarrow 0$, 
hence 
\begin{equation}
P_{F} \approx 1-|S_{\ell}|^{2} \rightarrow 1
          \;\;\;\;\;\;\;\;\;\; \mbox{for small} \;\; \ell.
\end{equation}
The last relation follows from the fact that for such a strong absorptive case 
as in heavy ion collisions, $S_{\ell}$ becomes 
essentially zero for small $\ell$. 
Since $P_{E}+P_{D} \rightarrow 0$ for small $\ell$, 
$P_{E}+P_{D}+P_{F} \rightarrow 1$ as expected earlier in Eq.~(1.4).
In the next subsection, we further study this point numerically.
 
 In passing, we remark that the procedure we have proposed can also be used 
to reduce the quantum mechanical Rutherford cross section to the classical one.
As is well known, this reduction has been given by using a set of 
semi-classical approximations~\cite{mott,ford}. 
The quantum mechanical Rutherford cross section has the well known form,
\begin{equation}
\frac{d\sigma_{c}}{d\Omega} = 
 \left| \frac{1}{2ik} \sum_{\ell} (2\ell +1)
          (e^{2i\eta_{\ell}}-1)P_{\ell}(\theta) \right|^{2},
\end{equation}
where $\eta_{\ell}$ is the Coulomb phase shift and $P_{\ell}$ is the
Legendre function. One of the approximations introduced in the 
reduction process is to ignore the term
$-\frac{1}{2ik}\sum_{\ell}(2\ell+1)P_{\ell}(\theta)$. 
This term gives rise to a divergent contribution at extremely
forward angles and we ignore it as is usually done (see Refs.~\cite{mott,ford}).
We then integrate Eq.~(2.10) over angles to obtain the total elastic cross
section expressed as a sum of the partial cross sections, which is in turn
converted to an integral over $\ell$. The resultant total elastic scattering 
cross section takes a very simple form, namely 
\begin{equation}
\sigma_{c}=\sum_{\ell} \sigma_{c;\ell}=\int \frac{d\sigma_{c}}{d\ell} d\ell 
     \;\;\; \mbox{with} \;\;\;
\frac{d\sigma_{c}}{d\ell}=\sigma_{c;\ell}=\frac{\pi}{k^{2}}(2\ell +1) 
    \approx \frac{\pi}{k^{2}}(2\ell).
\end{equation}
By inserting the last expression in Eq.~(2.11) into Eq.~(2.5),
we obtain the Rutherford differential cross section.  
                                         
 It appears that the procedure used for reducing the quantum mechanical 
Rutherford cross section to the classical one involves a contradictory
element; we first integrate the differential cross section over angle, 
but then recover it from the partial wave cross sections we obtained 
as a result of the angle integration. 
However, the procedure can be justified:
The quantum mechanical cross section Eq.~(2.10) is given as 
a coherent sum over $\ell$. 
As has been demonstrated in a number of semi-classical treatments of 
Eq.~(2.10)~\cite{mott,ford}, the dominant contribution to 
the differential cross section for a given scattering angle $\theta$ 
comes from the partial waves around $\ell=\ell_{\theta}$, where $\ell_{\theta}$
is related to $\theta$ by Eq.~(1.3). 
The contribution becomes $\delta$-function-like in the classical limit of 
$\hbar \rightarrow 0$. In the present procedure, we carry out an integration 
over $\theta$ first, but from what has been discussed above, it is seen
that the contribution from the angle $\theta$ is stored into 
the partial wave cross section of $\ell \approx \ell_{\theta}$.
It is thus justifiable to recover the differential cross section 
at an angle $\theta$ from the partial wave cross section for 
$\ell = \ell_{\theta}$. 
This procedure is most justified when the scattering is closest to
classical.

\subsection{Numerical Examples}

 Calculations of $P_{E}$, $P_{D}$, and $P_{F}$ are performed for
the $^{16}$O+$^{208}$Pb system with incident energy $E_{lab}$=90~MeV,
using the optical model potential as fixed in our previous study~\cite{kim1}. 
We present the results in Fig.~2, where the solid, (thick and thin) dotted, 
and (thick and thin) dashed curves
are the calculated values of $P_{E}$, $P_{D}$ and $P_{F}$, respectively. 
The experimental data for $P_{E}$ and $P_{D}$ are also 
plotted by the solid and open circles, respectively. 
As seen, the calculated $P_E$ reproduces the experimental $P_{E}$ very well. 
This is not the case, however, for $P_{D}$; the calculated $P_{D}$ is shifted
to the smaller $d$-region by about 0.05~fm 
as compared with the experimental data, 
particularly in the region of $d\geq 1.6$~fm. 
Thus, the comparison of the calculated $P_{D}$ with the data 
provides an additional test of the parameters  used in Ref.~\cite{kim1}.
In fact, this shift means the radius parameter $r_{D}=1.50$~fm used in 
Ref.~\cite{kim1} is too small to describe the data. 
We thus repeated the calculation with a larger  
radius parameter of $r_{D}=1.55$~fm. 
The calculated $P_{D}$ thus obtained is plotted by the thin dotted curve
shown in Fig.~2.
The fit to the data is improved. 
It is worth noting that the recalculated $P_{D}$ shifts toward 
the larger $d$-region by 0.05~fm, the same amount as the increase of 
the radius parameter $r_{D}$.  
This shows that the experimental $P_{D}$ provides 
a very sensitive test of the $r_{D}$-value. Based on the result obtained
above, we use the value $r_{D}$=1.55~fm in the 
$\chi^{2}$-analysis discussed in the next section.

 In the calculation of $P_{F}$ shown in Fig.~2, use is made of $r_{F}$=1.40~fm.
Thus the $P_{F}$ curve lies in much smaller $d$-region than the $P_{D}$.
We also observe that the slope of $P_{F}$ is much steeper than that of $P_{D}$. 
This reflects the fact that the diffuseness parameter
used for $W_{F}(r;E)$ ($a_{F}$=0.25~fm) is smaller than that of $W_{D}(r;E)$
($a_{D}=0.45$~fm). To show the effects of the $a_{F}$-value, we present by 
the thin dashed curve another $P_{F}$ calculated with $a_{F}$=0.45~fm.
It is seen that the slope of $P_{F}$  at large distances is almost the
same as that of $P_{D}$ with $a_{D}=0.45$~fm.
 
 Let us now turn to the sum $P_{E}+P_{D}+P_{F}$ shown by the dash-dotted 
curve. As expected, it stays
very close to unity, confirming that the relation Eq.~(1.4) is fairly well 
satisfied, within the accuracy of, say, 20 \%.  
The sum shows some oscillations around unity, 
which may be ascribed to quantum interference effects. 
The oscillation is also visible in the experimental $P_E$ values. 
Accumulation of more accurate data may enable us to test this 
explanation in a more detailed manner.
                           
 Presented in Fig.~3 are the $P_{E}$-values calculated for the incident 
energies considered in Fig.~1. 
Use is made of the optical potential determined from the $\chi^{2}$-analysis 
discussed in the next section.  
Since we use such a potential as determined from the $\chi^{2}$-fit, 
the calculated $P_{E}$ fit the data given in Fig.~1 very well 
and thus they forms a band very much similar to that seen in Fig.~1.

\section{$\chi^{2}$-Analyses}

 We have repeated simultaneous $\chi^{2}$-analyses as  
in Ref.~\cite{kim1} for the elastic scattering, DR, 
and fusion data for the $^{16}$O+$^{208}$Pb system. 
This is motivated for two reasons: 
The first is that the fusion data have been revised~\cite{mort}, after 
Ref.~\cite{kim1} was published. 
The second is that we are now able to test the calculations 
against the data for $P_{D}$. 
As described in the previous section, the value of the radius parameter 
$r_{D}$=1.50~fm used in Ref.~\cite{kim1} is too small to explain the data. 
A better $r_{D}$-value is $r_{D}=$1.55~fm. Other parameters
must be fixed with this more appropriate value of $r_{D}$. 
As in Ref.~\cite{kim1}, we utilize a dispersive type optical potential~\cite{maha}
\begin{equation}
U=U_{C}(r)-[V_{0}(r)+V(r;E)+iW(r;E)],
\end{equation}
where $U_{C}(r)$ is the Coulomb potential and $V_{0}(r)$ is
the energy independent Hartree-Fock part of the potential, 
while $V(r;E)+iW(r;E)$ is the polarization part of the potential~\cite{love} 
that originates from couplings to reaction channels.  
They are assumed to have volume-type fusion and surface-derivative-type DR
parts. 
Explicitly, $V_{0}(r)$, $V(r;E)$ and $W(r;E)$ are given, respectively, by
\begin{equation}
V_{0}(r)=V_{0}f(X_{0}),
\end{equation}
\begin{equation}
V(r;E) = V_{F}(r;E)+V_{D}(r;E)=
    V_{F}(E)f(X_{F})+4V_{D}(E)a_{D}\frac{df(X_{D})}{dR_{D}},
\end{equation}
\begin{equation}
W(r;E) = W_{F}(r;E)+W_{D}(r;E) = 
    W_{F}(E)f(X_{F})+4W_{D}(E)a_{D}\frac{df(X_{D})}{dR_{D}},
\vspace{2ex}
\end{equation}
where $f(X_{i})=[1+\mbox{exp}(X_{i})]^{-1}$ with $X_{i}=(r-R_{i})/a_{i}$
$({\it i}=0, \; D \; \mbox{or} \; F)$ is the usual Woods-Saxon function.
Use is made of the values used in Ref.~\cite{kim1} 
for the parameters of the bare potential
$V_{0}(r)$, and the geometrical parameters of $V(r;E)$ and $W(r;E)$ 
(except $r_{D}$ as discussed above);  $V_{0}$=60.4~MeV, $r_{0}$=1.176~fm, 
$a_{0}$=0.658~fm, $r_{F}$=1.40~fm, $a_{F}$=0.25~fm, $r_{D}$=1.55~fm, 
and $a_{D}$=0.45~fm. 
Once the geometrical parameters are fixed, the dispersion relation is 
reduced to a relation for the strength parameters $V_{i}(E)$ and 
$W_{i}(E)$ ($i = D$ and $F$). The relation now reads~\cite{maha} 
\begin {equation}
V_{i}(E)=V_{i}(E_{s}) + \frac {E-E_{s}}{\pi } P\int_{0}^{\infty} dE'
\frac {W_{i}(E')}{(E'-E_{s})(E'-E)},
\vspace{2ex}
\end {equation}
where $P$ stands for the principal value and $V_{i}(E_{s})$ is 
the potential value at a reference energy point $E=E_{s}$.

 As was done in Ref.~\cite{kim1}, we approximate the $E$-dependence of 
$W_{i}(E)$ just above the threshold energy $E_{0;i}$ (defined as
$W_{i}(E_{0;i})=0$) by a linear function of $E$. (See the forthcoming
Eqs.~(3.7) and (3.8).)
We then identify this threshold energy as that determined from the 
linear representation of the quantity $S_{i}(E)$ introduced 
by Stelson {\it et al.}~\cite{stel} as
\begin{equation}
S_{i} \equiv \sqrt{E\sigma_{i}(E)} \propto (E-E_{0;i}) 
        \;\;\;\;\;\;\;\;\;\ (i= D\; \mbox{or}\; F). 
\end{equation}
The threshold energies $E_{0;i}$ thus defined are essentially 
the threshold energies of the DR ($i= D$) and fusion ($i= F$) cross sections,
and it is plausible to identify the two threshold energies to be the same. 
The authors of Ref.~\cite{stel} considered the quantity $S_{i}$ only 
for the $i = F$ case, but we extend it to DR.
Originally, two threshold phenomena in the imaginary part of the optical 
potential and the fusion cross section data were found independently, but 
it was noticed later~\cite{uda3} that the two are very close to one another. 
Once we have separated the imaginary potential into the DR and fusion parts, 
it is physically plausible to require that the two thresholds should 
be the same. 
In Fig.~4, we present the $S_{i}$-values for $i= D$ and $F$.
There we find that $E_{0;D}=$73.0~MeV and $E_{0;F}$=76.0~MeV, which will 
be used later as the threshold energies of $W_{D}(E)$ and 
$W_{F}(E)$, respectively.
  
 In an attempt to determine the polarization potential, simultaneous
$\chi^{2}$-analyses were performed, treating all four strength parameters, 
$V_{D}$, $V_{F}$, $W_{D}$ and $W_{F}$ as the adjustable parameters. 
We took into account all the data~\cite{vide,valg} available 
for incident energies between $E_{lab}$=80~MeV and 104~MeV. 
We included the total DR and fusion cross sections in the analyses.

 The values of the parameters thus extracted are presented in Fig.~5 for 
$V_{D}$ and $W_{D}$ and in Fig.~6 for $V_{F}$ and $W_{F}$. 
Let us consider first the results for $V_{D}$ and $W_{D}$. 
A considerable fluctuation is seen in the values of $V_{D}$, 
but $W_{D}$ changes smoothly as a function of $E$. 
The fact that $W_{D}$ could be fixed as a smooth function of $E$ 
indicates that these values are reliable. 
There is a reason that $W_{D}$ can be determined rather 
unambiguously and becomes a smooth function of $E$. 
It is because $W_{D}$ is the dominant absorptive term in
the peripheral region. 
Therefore, the elastic scattering  cross section is quite sensitive 
to the value of $W_{D}$. 
This is not the case for $V_{D}$; at the strong absorption radius, where the 
elastic scattering cross section is sensitive to the real potential,
$V_{D}$ is generally much smaller than the bare potential $V_{0}(r)$, 
resulting in difficulty in determining $V_D$ unambiguously.  
The fluctuation seen in Fig.~5(a) may be understood to arise from 
this difficulty.

 The $W_{D}$-values determined from the $\chi^{2}$-analyses can be
well represented by the following function of $E$ (in units of MeV):
\begin{equation}
W_{D}(E) \; = \; \left\{ \begin{array}{lll}
                         0            & \mbox{for    $E \leq$73.0}    \\
                         0.015(E-73.0) & \mbox{for 73.0$< E \leq$92.5} \\
                         0.2925        & \mbox{for  $ E >$ 92.5}
                         \end{array}
                 \right.
\vspace{2ex}
\end{equation}
where $E_{0;D}$=73.0~MeV is used as extracted in Fig.~4.
The solid line shown in Fig.~5(b) is $W_{D}$ given by Eq.~(3.7).
The line fits the empirical values quite well.
                         
 Since a reliable value of $W_{D}$ is now available, one can calculate
$V_{D}$ by using the dispersion relation Eq.~(3.5). 
In doing this, we need to know one more parameter, {\it i.e.}, 
the value of $V_{D}$  at $E=E_{s}$. 
We may fix this $V_{D}(E_{s})$ by fitting the average of the resultant $V_{D}$
to that of the empirically determined $V_{D}$.  
The solid curve shown in Fig.~5(a) shows the $V_{D}$-values thus calculated. 
The $V_{D}(E_{s})$-value used is $V_{D}(E_{s})$=0.4~MeV at $E_{s}$=92.5~MeV.

 As seen in Fig.~6, $V_{F}$ and $W_{F}$ are both determined as fairly 
smooth functions of $E$. 
The $W_{F}$-values may be represented (in units of MeV) as
\begin{equation}
W_{F}(E) \; = \; \left  \{ \begin{array}{lll}
                            0                & \mbox{for $E\leq$76.0}    \\
                            0.32(E-76.0)     & \mbox{for 76.0$< E \leq$86.0} \\
                            3.2            & \mbox{for $E >$ 86.0 }.
                           \end{array}
                 \right.
\vspace{2ex}
\end{equation}
Again we took the threshold energy of $E_{0;F}$=76.0~MeV 
determined from $S_{F}$. The solid line shown in Fig.~6(b) represents
$W_F$ in Eq.~(3.8).
We then calculated the $V_{F}(E)$ by 
using the dispersion relation Eq.~(3.5) with $W_{F}$ given by Eq.~(3.8). 
The reference potential $V_{F}(E_{s})$ 
involved was chosen as $V_{F}(E_{s})$=3.50~MeV at $E_{s}$=86.0~MeV.
As shown by the solid curve in Fig.~6(a), 
the predicted $V_{F}$-values again agree reasonably well 
with the empirically determined values. 
                                
 We take  as our final potential parameters
$W_{D}$ and  $W_{F}$ given, respectively, by
Eqs.~(3.7) and (3.8), and also $V_{D}$ and $V_{F}$ generated from the
dispersion relation Eq.~(3.5). 
The potential with such parameters then fully satisfies 
the dispersion relation. 
Using such an optical potential, we calculated the final theoretical $P_{E}$,
$\sigma_{D}$, and $\sigma_{F}$ and presented them in Figs.~7 
and 8 in comparison with the experimental data.
As seen, all experimental $P_{E}$, $\sigma_{D}$ and $\sigma_{F}$  
are well reproduced by the calculations.

 We now wish to give some remarks on the polarization potential we obtained.
First, there is a remarkable difference in the the
energy dependences between the DR and fusion potentials. A very rapid
change is seen only in the fusion part of the potential. The slope of
$W_{F}(E)$ given by Eq.~(3.8) in the threshold region
is 0.32, while that in $W_{D}(E)$ given by Eq.~(3.7) is only
0.015. As a result, we see a significant energy variation 
of about 2~MeV in $V_{F}(E)$ in the interval of $\sim$ 10~MeV, but 
the change in $V_{D}$ is only 0.1~MeV in 
the energy range of $\sim$ 20~MeV. We may thus conclude that the threshold 
anomaly exists in the fusion part of the potential, but not in
the DR part.

\section{Concluding Remarks}
                    
 We have presented a simple method to calculate the DR and fusion
probabilities within the optical model by introducing two types of
imaginary potentials, DR and fusion. 
These probabilities are calculated by using the partial DR and 
fusion cross sections generated from the corresponding imaginary potentials
with the help of the classical relation 
between the orbital angular momentum $\ell$ and the scattering angle $\theta$. 
The probabilities thus calculated were shown to satisfy the condition that 
the value should be equal to or less than unity.
                                       
 Based on the expressions derived, numerical calculations of these probability
were performed. We found that the sum of the DR, fusion and elastic
probabilities stays close to unity. We also analyzed 
the angular distribution data of the inclusive DR cross section, 
demonstrating that the data provide useful information 
for determining the radius parameters of the DR potential. It was observed
that a very rapid energy variation (threshold anomaly) was 
in the fusion part of the potential, but it is hardly seen in the DR part,
particularly in the real part of the potential. 
                                                
 A simultaneous $\chi^{2}$-analysis of elastic scattering, DR and fusion 
cross sections for the $^{16}$O+$^{208}$Pb system 
at near-barrier energies were performed for determining 
the polarization part of the optical potential 
that satisfies the dispersion relation over all space. 
The potential thus determined is found to reproduce the data well.  
                                                        
 The authors wish to express their sincere thanks to Prof. W. R. Coker
for his kind reading of the manuscript and comments.
One of the authors (BTK) acknowledges the support by Korea Research 
Foundation (KRF-2000-DP0085).

\newpage

\setlength{\leftmargin}{6em}
\begin {center}
FIGURE CAPTIONS
\end {center}
 
\vspace {1ex}\hfill\break
\par
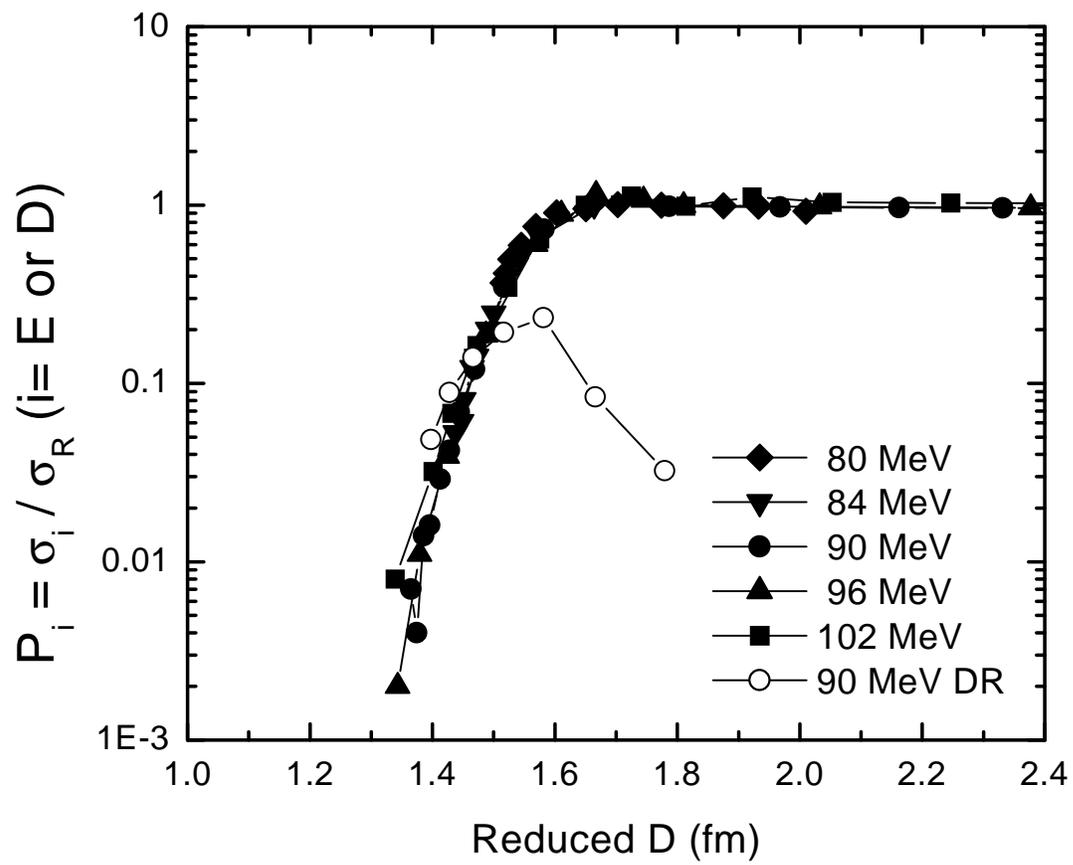
Fig.~1.~The experimental elastic probabilities, $P_{E}$, as a function of the 
reduced distance D for the $^{16}$O+$^{208}$Pb system 
at $E_{lab}=$~80, 84, 90, 96, and 102 MeV. 
The measured DR probabilities, $P_{D}$, are also plotted at $E_{lab}=$~90 MeV. 
The data are taken from Refs.~3 and 4. 

\vspace {1ex}\hfill\break
\par
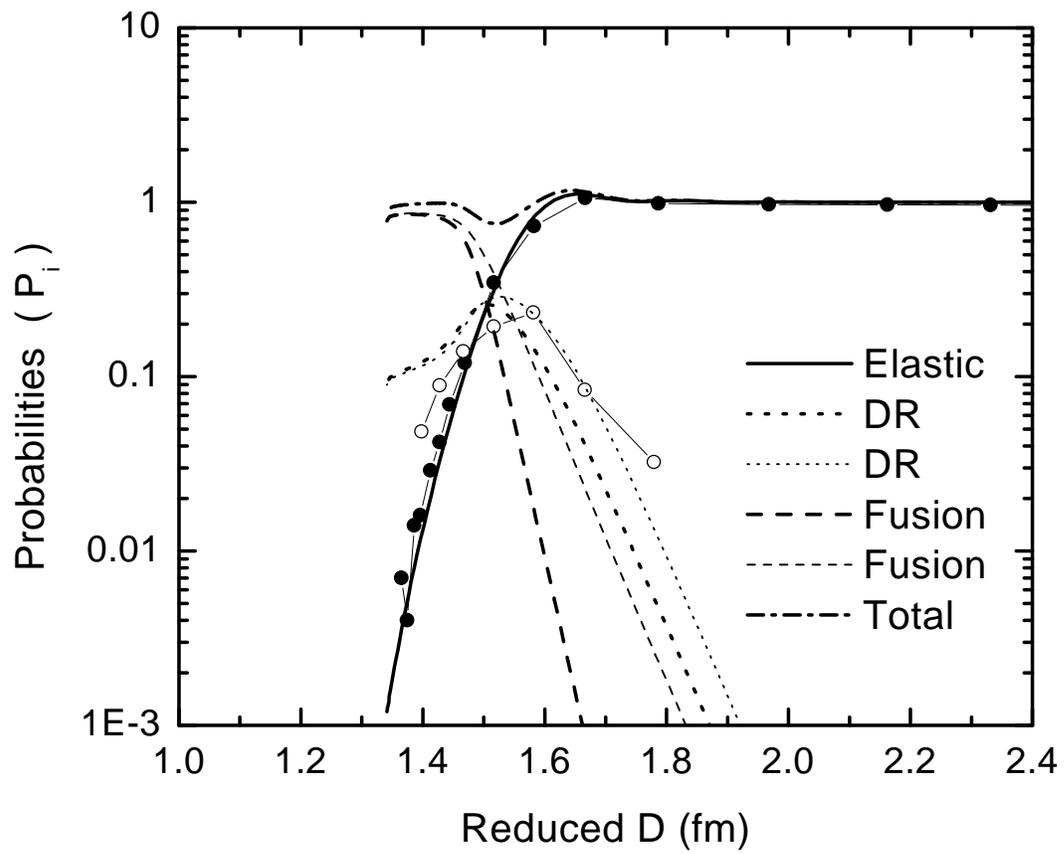
Fig.~2.~The optical model predictions of elastic, DR, and fusion probabilities 
as a function of the reduced distance D 
for the $^{16}$O+$^{208}$Pb system at $E_{lab}=$~90 MeV are shown
in comparison with the experimental ones.
The theoretical total probabilities are also shown. 
The thick dotted curve denotes $P_D$ calculated with 
the dispersive optical potential determined in Ref.~7.
The thin dotted curve represents $P_D$ calculated with the same potential
but with modified $r_{D}$ (=1.55 fm). The thick (thin) dashed curve denotes
$P_F$ calculated with $a_{F} = 0.25$~fm ($a_F =0.45$~fm). 

\vspace {1ex}\hfill\break
\par
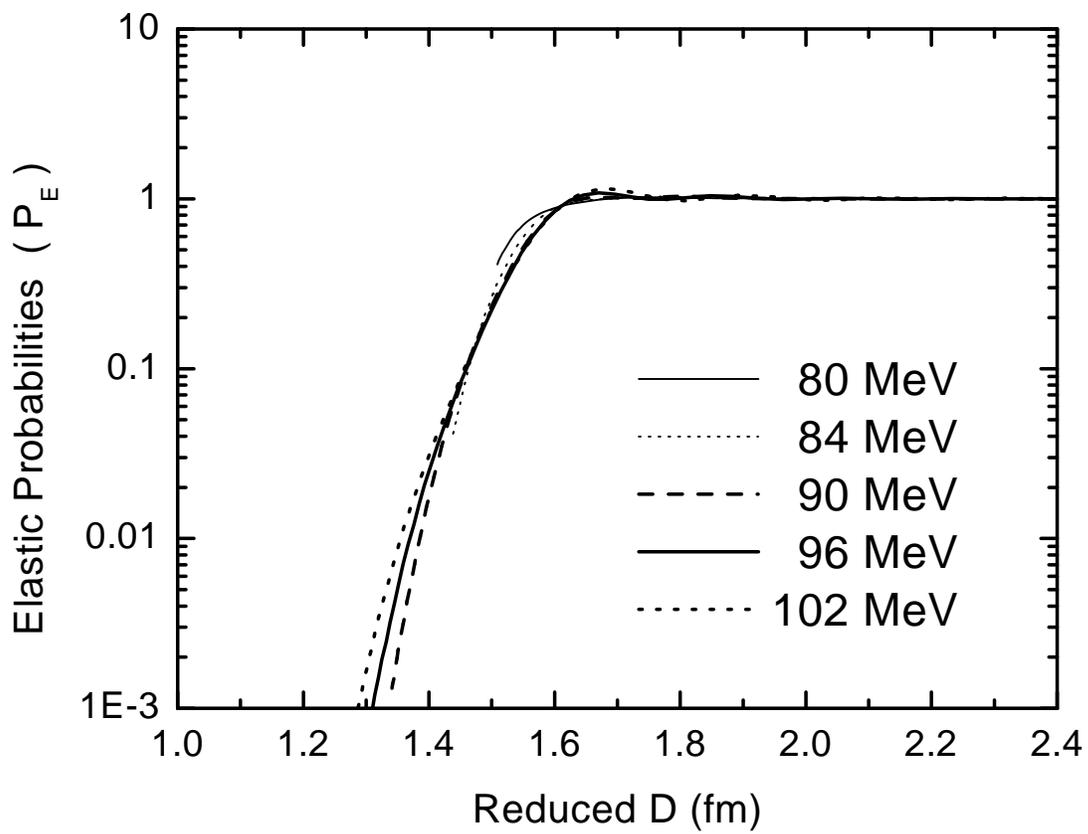
Fig.~3.~The results of calculations for elastic probabilities, 
$P_{E}$, as a function of 
the reduced distance D for the $^{16}$O+$^{208}$Pb system 
at $E_{lab}=$~80, 84, 90, 96, and 102 MeV, using the final  
fully dispersive optical potential.

\vspace {1ex}\hfill\break
\par
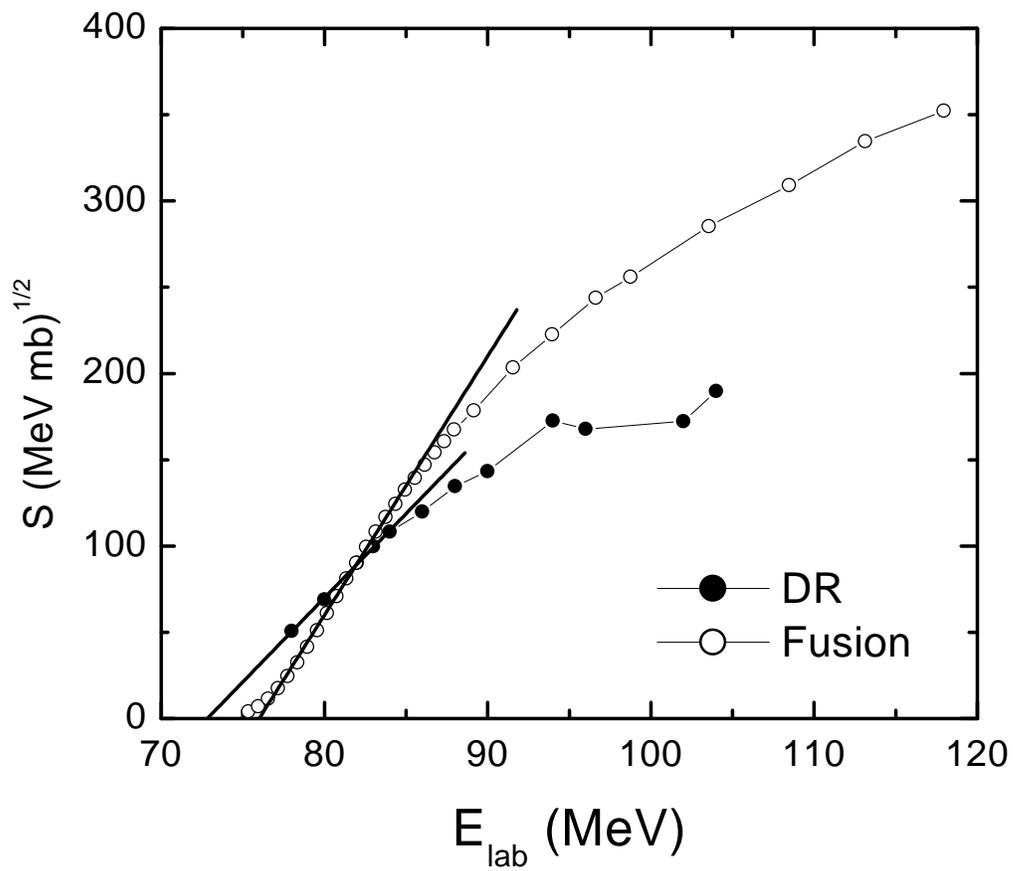
Fig.~4.~The Stelson plot of $S_{i}= \sqrt{E_{cm}\sigma_{i}}$ for 
direct reaction ($i= D$, solid circles) and 
fusion ($i= F$, open circles) cross sections. 
The straight lines are drawn to show the extraction of
the threshold energies. 
Thin lines connecting the circles are only to guide the eyes.

\vspace {1ex}\hfill\break
\par
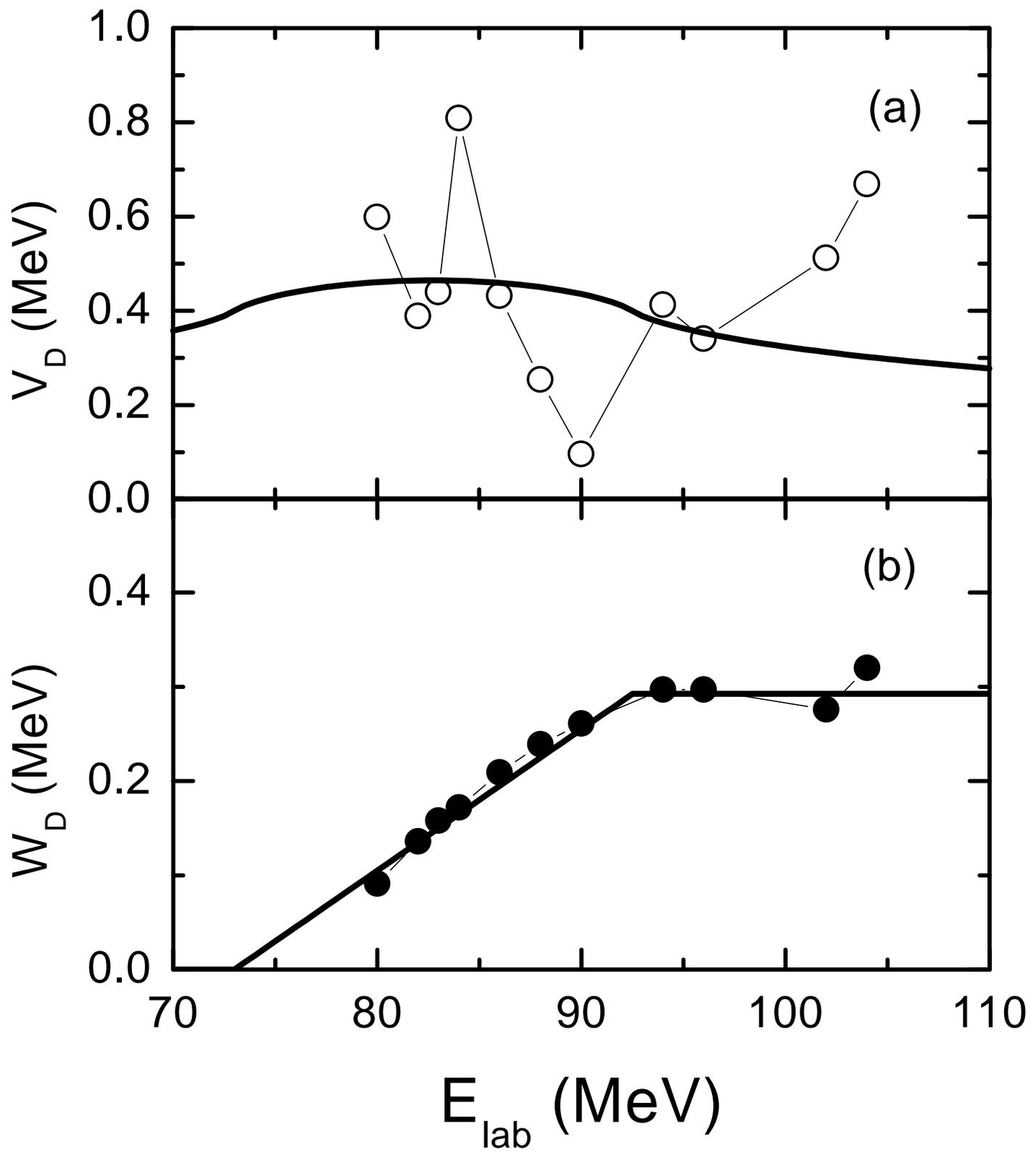
Fig.~5.~The strength parameters $V_{D}(E)$ and $W_{D}(E)$
for the direct reaction potential as  functions of $E$. 
The open and solid circles are the values extracted from 
the $\chi^{2}$-analyses. The solid lines denote
$W_{D}(E)$ and $V_{D}(E)$ calculated, respectively, 
from Eq.~(3.7) and from Eq.~(3.5) together with
Eq.~(3.7). The thin lines connecting the circles are 
only to guide the eyes.

\vspace {1ex}\hfill\break
\par
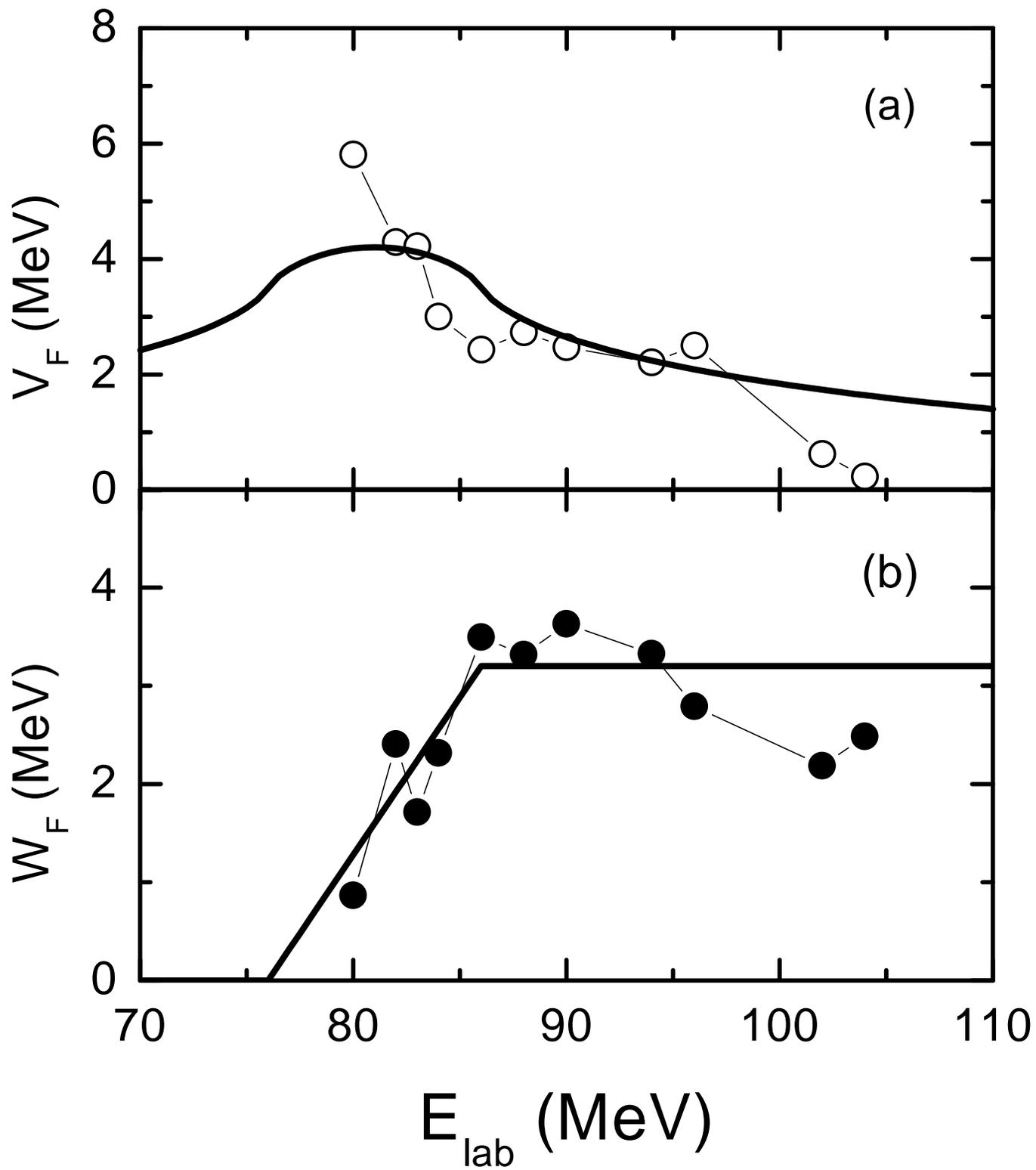
Fig.~6.~The strength parameters $V_{F}(E)$ and $W_{F}(E)$
for the fusion potential as functions of $E$. 
The open and solid circles are the values extracted from 
the $\chi^{2}$-analyses.  The solid lines denote  
$W_{F}(E)$ and $V_{F}(E)$ calculated, respectively,
with Eq.~(3.8) and Eq.~(3.5).
The thin lines are to guide the eyes.

\vspace {1ex}\hfill\break
\par
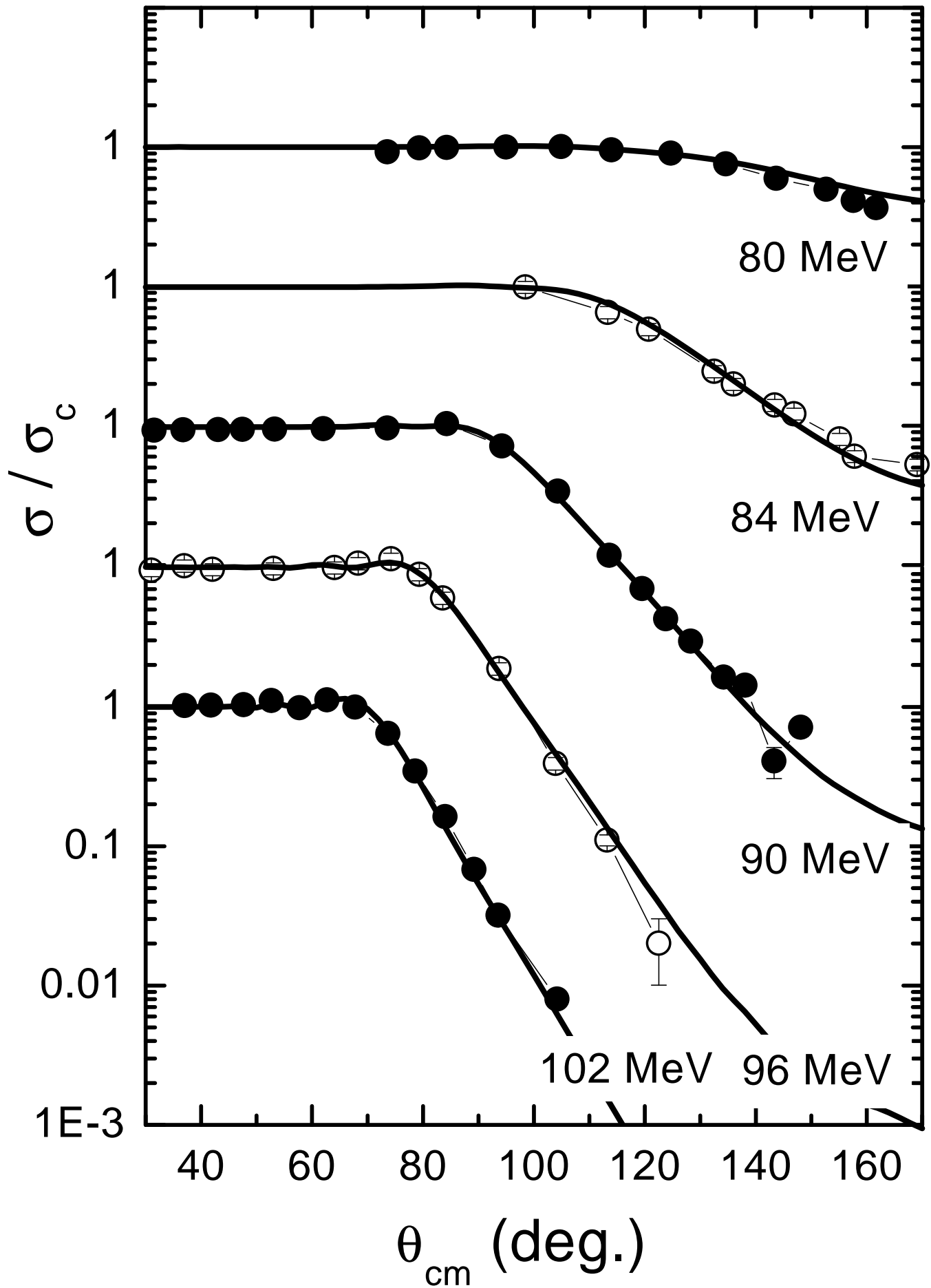
Fig.~7.~The ratios of the elastic scattering 
cross sections to Rutherford cross sections calculated with our final optical 
potential for the $^{16}$O+$^{208}$Pb system are shown in comparison with
the experimental data. 
The data are taken from Refs.~3 and 4.   

\vspace {1ex}\hfill\break
\par
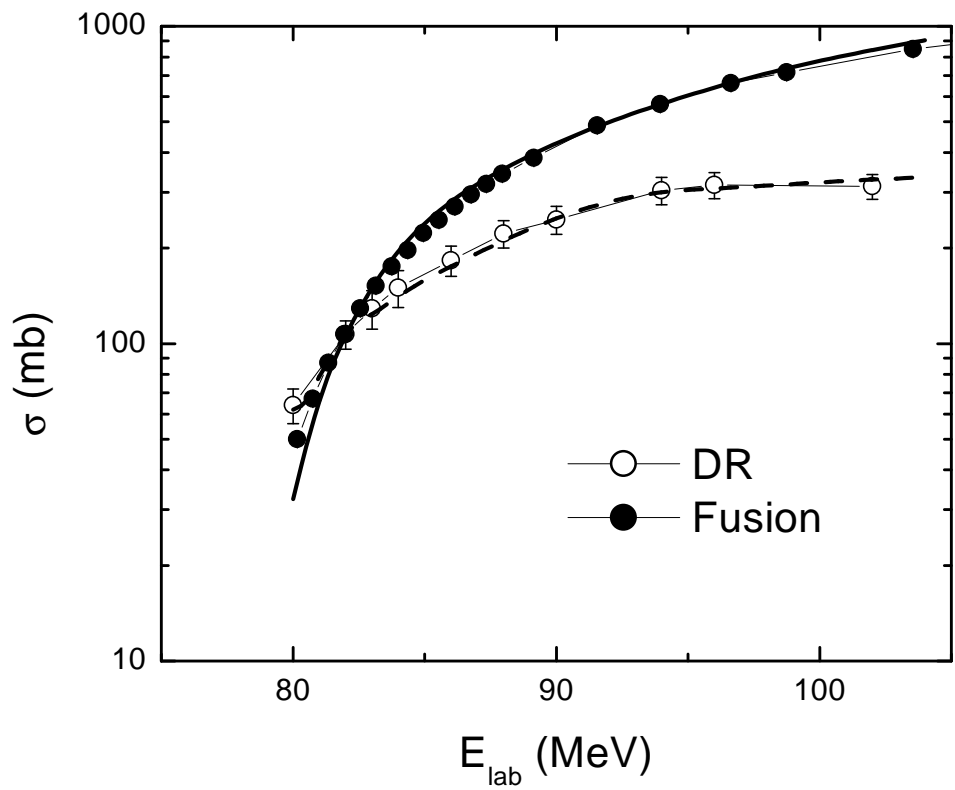
Fig.~8.~The direct reaction and fusion cross sections calculated with
our final optical potential for the
$^{16}$O+$^{208}$Pb system are shown in comparison with the experimental data. 
The direct reaction data are taken from Refs.~3 and 4, 
while the fusion ones from Ref. 8.


\newpage


\begin{references}

\baselineskip=22pt

\bibitem{bass} R. Bass, {\it Nuclear Reactions with Heavy Ions} 
               (Springer-Verlag, New York, 1980) 
\bibitem{sat1} G. R. Satchler, {\it Introduction to Nuclear Reactions}
               (John Wiley \& Sons, New York, 1980) p.41.
\bibitem{vide} F. Videvaek  {\it et al.}, Phys. Rev. C
               {\bf 15}, 954 (1977).
\bibitem{valg} E. Vulgaris,  L.  Gradzins, S. G .Steadman, and  R. Ledoux, 
               Phys. Rev. C {\bf 33}, 2017 (1986).
\bibitem{hong} S.-W. Hong, T. Udagawa, and T. Tamura, Nucl. Phys. 
               {\bf A491}, 492 (1989).
\bibitem{uda1} T. Udagawa, T. Tamura, and B. T. Kim, Phys.  Rev. C 
               {\bf 39}, 1840 (1989).
\bibitem{kim1} B. T. Kim, M. Naito, and T. Udagawa, Phys. Lett. B
               {\bf 237}, 19 (1990). 
\bibitem{mort} C. R. Morton, A. C. Berriman, M. Dasgupta, D. J. Hinde, 
               J. O. Newton, K. Hagino, and I. J. Thompson, Phys. Rev. C 
               {\bf 60}, 044608 (1999).
\bibitem{uda2} T. Udagawa and T. Tamura, Phys. Rev. C  {\bf 29}, 1922
               (1984); T. Udagawa, B. T. Kim, and T. Tamura, Phys. Rev. C 
               {\bf 32}, 124 (1985).
\bibitem{huss} M. S. Hussein, Phys. Rev. C {\bf 30}, 1962 (1984).
\bibitem{mott} N. F. Mott and H. S. W. Massey, {\it The Theory of Atomic
               Collisions}, (Oxford University Press, 1965) p.97-102.
\bibitem{ford} K. W. Ford and J. A. Wheeler, Ann. Phys. {\bf 7}, 287 (1959).
\bibitem{maha} C. Mahaux, H. Ngo, and G. R. Satchler, Nucl. Phys. 
               {\bf A449}, 354 (1986); Nucl. Phys. {\bf A456},
               134 (1986); M. A. Nagarajan, C. Mahaux, and G. R. Satchler, 
               Phys. Rev. Lett. {\bf 54}, 1136 (1985).
\bibitem{love} W. G. Love, T. Terasawa, and G. R. Satchler, Nucl. Phys.
               {\bf A291}, 183 (1977).
\bibitem{sat3} G. R. Satchler, M. A. Nagarajan, J. S. Lilley, and 
               I. J. Thompson, Ann. of Phys. {\bf 178}, 110 (1987).
\bibitem{stel} P. H. Stelson, Phys. Lett. B {\bf 205}, 190 (1988);
               P. H. Stelson, H. J. Kim, M. Beckerman, D. Shapira, and
               R. L. Robinson, Phys. Rev. C {\bf 41}, 1584 (1990). 
\bibitem{uda3} T. Udagawa, M. Naito, and B. T. Kim, Phys. Rev. C {\bf 45},
               876 (1992).
\end{references}
\end {document}